\documentclass[journal]{IEEEtran}
\usepackage{amsmath}
\usepackage{amssymb}
\usepackage{graphicx}
\usepackage{epsfig}
\usepackage{algorithm}

\begin{document}

\newtheorem{Thm}{Theorem}[section]
\newtheorem{Cor}[Thm]{Corollary}
\newtheorem{Lem}[Thm]{Lemma}
\newtheorem{Prop}[Thm]{Proposition}
\newtheorem{Def}[Thm]{Definition}
\newtheorem{Rem}[Thm]{Remark}
\newtheorem{Assump}[Thm]{Assumption}
\newtheorem{Exam}[Thm]{Example}
%\numberwithin{equation}{subsection}
 %\thispagestyle{empty}
% \baselineskip 0.22in

%\title{\textbf{Theoretical Bound for Guaranteed Performance of Thresholding Methods for Signal Recovery}}
 \title{\textbf{Improved RIP-Based Bounds for Guaranteed Performance of two Compressed Sensing Algorithms}
 \thanks{The research was supported  by the Natural Science Foundation of China (NSFC) under the grant 12071307, and partially supported with grants 61571384, 61731018 and 11771003.} }
  % and the Leading Talents of Guang Dong Province Program [Grant %00201510]} }

%\title{\textbf{Improved RIP Bound for IHT, HTP and CoSaMP Algorithms for Signal Recovery}}

\author{Yun-Bin Zhao \emph{Member, IEEE}\thanks{Yun-Bin Zhao is with the Shenzhen Research Institute of Big Data, Chinese University of Hong Kong, Shenzhen, Guangdong, China (Email: yunbinzhao@cuhk.edu.cn). On leave from the University  of Birmingham, Birmingham B15 2TT,  United Kingdom (Email: y.zhao.2@bham.ac.uk).} ,   Zhi-Quan Luo \emph{Fellow, IEEE }\thanks{Zhi-Quan Luo is with the Shenzhen Research Institute of Big Data,  Chinese University of Hong Kong, Shenzhen, Guangdong, China
(Email: luozq@cuhk.edu.cn). }}

\date{ }

\maketitle

\begin{abstract} Iterative hard thresholding (IHT) and compressive sampling matching pursuit (CoSaMP) are two types of mainstream compressed sensing algorithms using hard thresholding operators  for signal recovery and approximation. The guaranteed performance for signal recovery via these algorithms has mainly been analyzed under the condition that the restricted isometry constant of a sensing matrix, denoted by $ \delta_K$ (where $K$ is an integer number), is smaller than a certain threshold value in the interval $(0,1).$ The condition $ \delta_{K}< \delta^*$ for some constant $ \delta^* \leq  1 $ ensuring the success of signal recovery with a specific algorithm  is called the restricted-isometry-property-based (RIP-based) bound for guaranteed performance of the algorithm.  At the moment, the best known RIP-based bound for the guaranteed recovery of $k$-sparse signals via IHT  is $\delta_{3k}< 1/\sqrt{3}\approx 0.5774,$  and  the bound for guaranteed recovery via CoSaMP is $\delta_{4k} < 0.4782. $  A fundamental question in this area is whether such theoretical results can be further improved.  The purpose of this paper is to affirmatively answer this question and rigorously show that the RIP-based bounds for guaranteed performance of IHT can be significantly improved to $ \delta_{3k} < (\sqrt{5}-1)/2 \approx 0.618, $  and the bound for CoSaMP can be improved and pushed to $ \delta_{4k}< 0.5102. $  These improvements are  achieved  through a deep property of the hard thresholding operator.

\end{abstract}

%\textbf{Key words:}   Iterative hard thresholding, compressive sampling matching pursuit,  subspace pursuit, restricted isometry property,  compressed sensing, signal recovery.


 % \textbf{AMS subject classifications:} 90C30, 90C05, 90C25, 65F10, 94A12, 15A29

%\newpage

\section{Introduction} \label{section1}

 One of the important tasks in signal processing is to  recover (reconstruct) an unknown signal from the linear and nonadaptive measurements acquired for the signal. The sparse or compressible signals arise in many scenarios especially when the signal is represented in certain transformed domains or over redundant bases \cite{B19}--\cite{M15}. The compressed sensing  algorithms were developed for signal recovery when the signal is sparse or can be sparsely approximated \cite{D06}--\cite{CT06}. The recovery of a sparse signal or the significant information of the signal usually amounts to solving a   sparse optimization model,  and the numerical methods for solving such a model  are often called compressed sensing algorithms (see, e.g., \cite{CT05, E10, EK12, FR13, Z18}).  Denote by $\|z\|_0$  the `$\ell_0$-norm'  counting the number of nonzero entries of  the $n$-dimensional vector $z \in \mathbb{R}^n.$  Let $ A $ be an $m \times n$ sensing  matrix with $m<n. $ The typical model for sparse signal recovery can be formulated as the   $\ell_0$-minimization problem
$$ \min\{\|z\|_0:  \| Az-y\|_2\leq  \epsilon \}, $$  where $\epsilon \geq 0$ is a given parameter and $ y: =Ax +\nu   $ are the measurements of the target signal $x \in \mathbb{R}^n  $ with measurement errors $\nu \in \mathbb{R}^m$ bounded as $\|\nu\|_2 \leq  \epsilon .$   The model above aims at finding the sparsest data $z^*$ that can best fit the linear measurements of $ x$ and thus under a suitable assumption the recovery $ z^*= x$ can be achieved. In  many practical situations, however,  one is interested in recovering only  the significant information of a signal which usually is interpreted as a small number of the largest absolute coefficients over the redundant bases of the signal (such as the redundant wavelet bases of a natural image).  Based on this consideration, the sparse recovery model can be formulated as the following minimization problem with a sparsity constraint:
\begin{equation} \label{L0} \min_{z} \{\|Az-y\|_2^2:  ~ \|z\|_0 \leq k\},\end{equation}
  where $k$ is a given integer number, the estimation of the sparsity level of the signal.    The purpose of the model (\ref{L0}) is to find the $k$-term approximation of the target signal such that the selected $k$ terms can best fit the acquired measurements compared to other $k$ terms. The model (\ref{L0}) is not only an essential model for sparse signal recovery to which several  compressed sensing algorithms have been developed (see, e.g., \cite {E10, EK12}, \cite{FR13}--\cite{NNW17}), but also an important model closely related to the low-rank matrix recovery \cite{CP11}--\cite{FS19}, variable selections in statistics \cite{BKM16}--\cite{NT20}, and sparse optimization and its various applications \cite{Z18}, \cite{BE13}--\cite{W19}.

 For the  model  (\ref{L0}), the basic algorithm using the hard thresholding operator is called the iterative hard thresholding (IHT)  \cite{BD08}--\cite{FR08} which admits several modifications such as the hard thresholding pursuit (HTP) \cite{F11},  the IHT with a fixed stepsize \cite{GK09}, the normalized iterative hard thresholding (NIHT) \cite{BD10},  the graded IHT  \cite{B14, BFH16}, and the recent Newton-step-based hard thresholding algorithms \cite{MZ20, ZXQ19}.   The more sophisticated methods using the hard thresholding operator include the well known compressive sampling matching pursuit (CoSaMP) \cite{NT09}  and subspace pursuit (SP) \cite{DM09}. Recent study of SP can be found in such references as \cite{GE12}-\cite{SXL14}.  The study in this paper is focused on the analysis of the  IHT and CoSaMP, two well known compressed sensing algorithms. The purpose  is to achieve remarkable improvement on the existing theoretical results concerning the guaranteed success in signal recovery/approximation with these algorithms.

 To describe the IHT and CoSaMP algorithms, let us first introduce a few notations.  We use $\mathbb{R}^n$ to denote  the $n$-dimensional Euclidean space and all vectors are understood as column vectors unless otherwise specified.  Given
a vector $z \in \mathbb{R}^n,$ the operator  $ \mathcal{H}_k(z) \in  \mathbb{R}^n $  called the hard thresholding operator  retains the $k$ largest absolute entries of $z$ and sets
other entries to zeros.  We use $L_k(z)$ to denote the index set of the  $k$ largest absolute entries of the vector $z,$ and we  use ${\rm supp} (z) = \{i: z_i\not=0\}$  to denote the support of the vector $z$, i.e.,  the index set of nonzero entries of $z.$
For a given vector $z\in \mathbb{R}^n  $ and matrix $A,$ the symbol $z^T$ and $A^T$ denote the transpose of $z$ and $A.$    Throughout the paper, a vector $x$  is said to be $k$-sparse if $\|x\|_0\leq k. $

The IHT \cite{BD08}--\cite{FR08} is a simple iterative scheme for the model (\ref{L0}), and is stated as Algorithm 1 below.
\begin{algorithm} \label{alg1}
\caption{ Iterative Hard Thresholding (IHT)} Input the  measurement matrix $A,$ measurement vector $y,$ and sparsity level $k.$ Perform the following steps:
\begin{itemize}
\item[S1]  Choose an initial $k$-sparse vector $x^0,$ typically $x^0 = 0;$

\item[S2] Repeat
$$x^{p+1}= {\cal H}_k( x^p+ A^T (y- A x^p))  $$ until a stopping criterion is met.

\end{itemize}

Output: the $k$-sparse vector $ \hat{x} . $
\end{algorithm}

 More efficient algorithms than the IHT can be obtained by integrating  an orthogonal projection into the algorithm (also called a pursuit step) (see, e.g., \cite{FR13, F11}).   Using both hard thresholding operator and orthogonal projection, the next algorithm (Algorithm 2) is referred to as compressive sampling matching pursuit (CoSaMP)  which was introduced by Needell and Tropp \cite{NT09}. The CoSaMP was closely related to an earlier greedy method called regularized orthogonal matching pursuit proposed by Needell and Vershynin \cite{NV09, NV10}.
\begin{algorithm} \label{alg2}
\caption{Compressive Sampling Matching Pursuit (CoSaMP)}
Input the measurement matrix A, measurement vector $y,$ and sparsity level $k$. Perform the steps below:

  \begin{itemize} \item[S1] Choose an  initial $k$-sparse vector $x^0,$ typically $x^0 = 0; $

\item[S2]   Repeat
\begin{align}
U^{p+1} &= \textrm{supp}(x^p) \cup  L_{2k}(A^T
(y - Ax^p)) ,   \tag{CP1} \\
 z^{p+1}& = \textrm{arg}\min_{
z\in \mathbb{R}^n} \{\|y - Az\|_2:  ~ \textrm{supp}(z) \subseteq U^{p+1}\} ,
    \tag{CP2} \\
x^{p+1} &= {\cal H} _k(z^{p+1})   \tag{CP3}
\end{align}
until a stopping criterion is met.

\end{itemize}

Output: the $k$-sparse vector  $\hat{x}. $
\end{algorithm}
The step (CP2) in CoSaMP is an orthogonal projection which seeks a vector that best fits the measurements over the prescribed support. As pointed out in \cite{Z20, ZL20}, the orthogonal projection may generally stabilize or speed up the IHT framework.

%The subspace pursuit (SP) (i.e., Algorithm 3)  introduced by Dai and Milenkovic %\cite{DM09} admits a similar structure to CoSaMP but with a different selection of the %index set  for the orthogonal projection which is performed twice in each loop instead of % only once in CoSaMP.

%\begin{algorithm} \label{alg3}
%\caption{Subspace Pursuit (SP)}
%Given $A, y, k$ and the initial iterate $x^0 =0$ with $ S^0= \textrm{supp}(x^0) = %\emptyset. $
%At the $k$-th  step, set $ S^p= \textrm{supp}(x^p)$ and perform the following steps to %generate the next iterate $x^{p+1}: $
%\begin{align*}   \Lambda^{p+1}  & = S^p \cup L_k(A^T (y- Ax^p)),    \\
% z^{p+1} &  = \arg\min_{z\in \mathbb{R}^n}  \{\| y- Az\|_2:   ~ \textrm{supp}(z) %\subseteq \Lambda^{p+1} \},   \\
%S^{p+1} & = L_k(z^{p+1})   , \\
%x^{p+1} & = \arg\min_{z\in \mathbb{R}^n}  \{\| y-Az\|: ~ \textrm{supp} (z) \subseteq %S^{p+1} \}.
%\end{align*}
 % Repeat the above steps until a certain stopping criterion is met.
%\end{algorithm}

The analyses for the guaranteed performance (including stability and convergence) of these algorithms were carried out widely in terms of the restricted isometry property (RIP) of the sensing matrix. The RIP and the associated restricted isometry constant (RIC) of order $K,$ denoted by $ \delta_{K},$ were first introduced by Cand\`es and Tao \cite{CT05, CT06}. The RIP tool is quite natural for the analysis of various compressed sensing algorithms.
The IHT  for compressive
sensing  was initiated by Blumensath and Davies in \cite{BD08} and was shown convergent under the condition $\delta_{3k}< 1/\sqrt{32}.  $  Stability and guaranteed performance for this method were established in \cite{BD09}
  under the condition $\delta_{3k}< 1/\sqrt{8}, $ which   is still rather restrictive.  This result was improved to $ \delta_{3k }< 1/\sqrt{3} \approx 0.5774$
  by Foucaut  in \cite{F11} (see also in \cite{FR13}). This bound remains the best bound for the
  hard thresholding pursuit (HTP) algorithm which is a simple  combination of IHT and orthogonal projection \cite{F11}. In this paper,  we will show that the current RIP-based bound for IHT is definitely not tight, and it can be improved to $ \delta_{3k }< (\sqrt{5}-1)/2 \approx 0.618.$ Certain evidences point to the conjecture that this new bound is optimal, i.e., the tightest one.

In \cite{NT09},  some theoretical results (stability and robustness) for CoSaMP
 were established  under the condition $ \delta_{4k} \leq 0.1.$ (Their proof actually implies that their results are valid under the bound
$ \delta_{4k} < 0.17157.$) This initial result was significantly improved to $ \delta_{4k} < 0.4782  $ by Foucart and Rauhut in \cite{FR13}.   In this paper, we will further improve this result to $\delta_{4k} < 0.5102. $ As seen later, such an improvement is far from being trivial and is achieved by establishing a deep property of the hard thresholding operator.

%The initial analysis for the SP algorithm was carried out by  Dai and Milenkovic %\cite{DM09} who   showed that the SP   converges under the condition $ \delta_{3k}   < %0.205. $   The RIP-based performance analysis of SP has also been done by other %researchers. The relatively earlier results  were established by Giryes and Elad %\cite{GE12} who showed that the stable signal recovery via SP can be ensured under the %condition $ \delta_{3k} < 0.139, $ and by Chang and Wu \cite{CW14} who established their %results under the condition  $ \delta_{3k}   < 0.2412. $  Recent remarkable improvements %of  these results have been achieved in \cite{LBJ13, SXL14}.  In \cite{LBJ13}, Lee, %Bresler and Junge proved that the  bound for SP can be improved to  $ \delta_{3k} < %0.325, $ and in \cite{SXL14},  Song, Xia and Liu showed that this bound can be %significantly improved to $ \delta_{3k}   < 0.4859.$    In this paper, we will further %improve these existing results for SP and prove that the signal recovery via SP can be %guaranteed under the condition  $ \delta_{3k}   < 0.5108 . $  We show such a result for a %more general algorithmic framework called general subspace pursuit (GSP) which %encompasses the SP as a special case.  The bound  $ \delta_{3k}   < 0.5108   $ is the %first convergent result rigorously shown for GSP in this paper.

The main contribution of the paper is summarized in the table below:

\vskip 0.1in

\begin{tabular}{|c||c|c|}
%{|p{1.5cm}| |p{2.5cm}|p{2.5cm}|}
% \hline
%\multicolumn{3}{|c|}{Improved Bounds for Guaranteed Performance} \\
 \hline
 ~Algorithms~     &  ~~ Existing results ~~   &  ~~New results~~      \\
 \hline
  IHT    &  $ \delta_{3k} < 0.5774 $   &    $ \delta_{3k} < 0.618 $ \\
  \hline
 CoSaMP &  $ \delta_{4k} < 0.4782  $  &  $ \delta_{4k} < 0.5102 $  \\
 %\hline
% GSP    &                             &    $ \delta_{4k} < 0.5108 $\\
% \hline
% SP     &  $ \delta_{4k} < 0.4859 $   &   $ \delta_{4k} < 0.5108 $   \\
 \hline
\end{tabular}\\

It is worth mentioning that  an open question for IHT and  CoSaMP remains standing: What is the optimal (i.e., the tightest) RIP-based bound for the algorithm?
Any improvement on RIP-based bounds for these algorithms moves closer to the unknown optimal bound which clearly exists in the interval (0,1) for every individual compressed sensing algorithm.   While it remains unclear at the moment whether the new results established in this paper for IHT and CoSaMP  are optimal or not, from the analysis in this paper it seems that  the room for a further improvement of our results is somewhat limited.

 The paper is organized as follows. In section \ref{section2}, we present a basic and deep property of the hard thresholding operator and use this property to show an improved RIP-bound for the guaranteed performance of signal recovery via IHT. In section \ref{section3}, we show an improved RIP bound for CoSaMP.

\section{Improved RIP bound for IHT} \label{section2}
 For a given set $ S \subseteq \{1,2,\dots, n\}, $  $|S|$ denotes  the cardinality of $S,$ and $\overline{S}=  \{1, 2,\dots, n\} \backslash S $  denotes the complement set of $S. $ The set difference of $S $ and  $ U$ is denoted by $S\backslash U = \{i:  i \in S, i\notin U\}.$  Given $S\subseteq \{1, \dots, n\} $ and a vector $x\in \mathbb{R}^n,$ the vector $x_S\in \mathbb{R}^n $ is obtained by retaining the entries of $x$ indexed by $S$ and zeroing out other components of $x.$

To establish improved convergence results  for the  algorithms IHT and  CoSaMP, we need to characterize the deep property of the hard thresholding operator $ {\cal H}_k (\cdot).$  Such a property will lead to the improved RIP-based bounds that guarantee the success of signal recovery with IHT and CoSaMP.

\vskip 0.1in

\begin{Lem} \label{Lem-01}  For any vector $z \in \mathbb{R}^n $ and  any $k$-sparse vector $x\in \mathbb{R}^n,  $    one has
\begin{equation} \label{ERR01}  \| ( x -      {\cal H}_k(z))_{S\backslash S^*} \|_2
     \leq   \|( x-z )_{ S \backslash S^*}\|_2
     +  \|( x -   z)_{S^*\backslash S} \|_2 ,
 \end{equation}
where  $S=  \textrm{supp} (x)  $ and  $   S^* = \textrm{supp} ( {\cal H}_k(z) ). $
\end{Lem}

\vskip 0.1in

\emph{Proof.}   By the definition of $ {\cal H}_k(\cdot)  ,$  we immediately see that $ \| z  - {\cal H}_k(z) \|_2^2 \leq \| z - d\|_2^2 $   for any  $k$-sparse vector  $d.$    In particular, setting  $d =  z_S ,$ where $S= \textrm{supp} (x), $ yields
  $$  \| z  - {\cal H}_k(z) \|_2^2 \leq \| z - z_S \|_2^2 =\| z _{\overline{S}} \|_2^2 =\| (z-x)_{  \overline{S}}\|_2^2 ,  $$ where the last equality follows from $ x_{\overline{S}}=0.$ Note that
  \begin{align*}   \| z  &   - {\cal H}_k(z) \|_2^2 = \| (z -x) +( x  - {\cal H}_k(z) )\|_2^2 \\
   &   = \|z  - x   \|^2_2  +  \|  x -  {\cal H}_k(z)    \|^2_2   -2 (x- {\cal H}_k(z)   )^T ( x-z ) .
  \end{align*}
   Therefore,
 \begin{equation} \label{III01} \|  x -   {\cal H}_k(z) \|^2_2 \leq  - \|(z -x )_S\|_2^2 +  2 (x- {\cal H}_k(z) )^T ( x-z) .   \end{equation}
Note that $ \textrm{supp} ( x-  {\cal H}_k(z) ) \subseteq S\cup S^* $  which can be decomposed into three disjoint sets $S\backslash S^*, S^*\backslash S$ and $S^* \cap S. $ We also note  that $  ({\cal H}_k(z))_i = z_i $ for every $ i \in S^*,$ and thus   $({\cal H}_k (z))_{S^*\setminus S} = z_{S^*\setminus S}$ and $({\cal H}_k(z))_{ S\cap  S^* } = z_{S^*\cap S}. $ The left-hand side of  (\ref{III01}) can be written as
\begin{align*}     \|  x     -         {\cal H}_k(z) \|^2_2    & =    \| [x -   {\cal H}_k(z)]_{S\backslash S^*} \|^2_2  +  \| [ x -   {\cal H}_k(z)]_{S^*\backslash S} \|^2_2 \\
 & ~~~ + \|  [ x -   {\cal H}_k(z)]_{ S^* \cap S} \|^2_2 \\
    & = \| [x -   {\cal H}_k(z)]_{S\backslash S^*} \|^2_2 +  \|(  x -   z)_{S^*\backslash S} \|^2_2  \\
  & ~~~  + \| ( x -  z)_{S^* \cap S} \|^2_2.
\end{align*}
  The right-hand side  of (\ref{III01}) is bounded as
     \begin{align*}
  &  -    \|      (z -x)_S\|_2^2+ 2 (x   -      {\cal H}_k(z) )^T ( x-z) \\
 & = - \|(z -x )_S\|_2^2+  2 [ (x- {\cal H}_k(z) )_{  S \backslash S^* }]^T  (x-z ) _{  S \backslash S^*  } \\
   & ~~~~ +  2 \|(  x -   z)_{S^*\backslash S} \|^2_2 +  2\| (x -  z)_{S^* \cap S} \|^2_2  \\
&  \leq  - \|(z -x)_{S \backslash S^*}\|_2^2+  2  \|[x- {\cal H}_k(z)]_{S \backslash S^* }\|_2  \|(x-z) _{  S \backslash S^*  } \|_2 \\
 &  ~~~~  +  2 \|( x -  z)_{S^*\backslash S} \|^2_2 + \| (x -  z)_{S^* \cap S} \|^2_2 .
\end{align*}
  Therefore, by substituting  the above two relations into (\ref{III01}) and cancelling and rearranging terms, we obtain that
\begin{align*}     \| [ x -      {\cal H}_k(z) ]_{S\backslash S^*}  &  \|^2_2
   \leq  - \|(z-x )_{S\backslash S^*}\|_2^2 +    \|(  x -   z)_{S^*\backslash S} \|^2_2 \\
 &   + 2\| (x- {\cal H}_k(z) )_{  S \backslash S^* }\|_2 \|  ( x-z ) _{  S \backslash S^* } \|_2.
\end{align*}
Thus  $  \| ( x -   {\cal H}_k(z))_{S\backslash S^*} \|_2 $ is  smaller than or equal to the largest real root of the  quadratic equation
\begin{align*}  Q(r):     =   &  r^2     -  2 r  \|( z -x )_{ S \backslash S^*}\|_2   +  \|(x-z )_{S\backslash S^*}\|_2^2  \\
&    -    \|(  x -   z)_{S^*\backslash S} \|^2_2   =0,
\end{align*}
to which the largest real root is given by
$$   r^*   =
    \|( x- z )_{ S \backslash S^*}\|_2+  \|(  x -   z)_{S^*\backslash S} \|_2.
$$
 Thus we immediately obtain the inequality (\ref{ERR01}).  \hfill $ \Box $

\vskip 0.1in

The next useful result is key to our later analysis.

\vskip 0.1in

\begin{Lem} \label{HHB}  For any vector $z\in \mathbb{R}^n$ and for any $k$-sparse vector $x\in \mathbb{R}^n$  (i.e., $\|x\|_0\leq k$), one has
\begin{equation} \label{TEB} \|x- {\cal H}_k(z)\|_2\leq \frac{\sqrt{5}+1}{2} \|(x-z)_{S\cup S^*}\|_2 ,
\end{equation}
where $  S= \textrm{supp} (x)$ and $S^*  = {\rm supp}  ({\cal H}_k(z)) .$
\end{Lem}

\vskip 0.1in

{\it Proof. }  By Lemma \ref{Lem-01},
 $ \|( x-{\cal H}_k(z))_{S\backslash  S^*}\|_2\leq \Delta_1+\Delta_2, $
where $ \Delta_1$ and $ \Delta_2$ are defined as
$$ \Delta_1= \| (x-z)_{S^*\backslash S} \|_2,  ~~ \Delta_2 = \| (x-z)_{S\backslash S^*} \|_2. $$
Thus,
\begin{align}  \label{III02}   \| x  &   -{\cal H}_k(z)\|_2^2=   \| (x     -{\cal H}_k)_{S\cup S^*} \|_2^2  \nonumber \\
& =  \|( x-{\cal H}_k(z))_{S^*}\|_2^2+\|( x-{\cal H}_k(z))_{S\backslash  S^*}\|_2^2 \nonumber \\
  & \leq \|( x-{\cal H}_k(z))_{S^*}\|_2^2   + \left ( \Delta_1 + \Delta_2\right)^2 \nonumber  \\
 & = \|( x-{\cal H}_k(z)) _{S^*\backslash S}\|_2^2 + \|( x-{\cal H}_k(z))_{S^*\cap S}\|_2^2  \nonumber \\
 & ~~~ +  \left ( \Delta_1 + \Delta_2\right)^2  \nonumber \\
 & =  \|( x-z)_{S^*\backslash S}\|_2^2 + \|( x-z)_{S^*\cap S}\|_2^2  \nonumber \\
  & ~~~ +  \left ( \Delta_1 + \Delta_2\right)^2.
\end{align}
Let $ C := \| (x-z)_{S^* \cup S} \|_2.$  We see that
$ C^2 = \| (x-z)_{S^* \cap S} \|_2^2+ \Delta_1^2+ \Delta_2^2  , $
and thus
$$ \| (x-z)_{S^* \backslash S} \|_2^2+ \| (x-z)_{S^* \cap S} \|_2^2 = C^2-\Delta_2 ^2 . $$
Substituting this relation into (\ref{III02})  yields
  $$   \| x-{\cal H}_k(z)\|_2^2   \leq C^2 + \Delta_1^2 + 2 \Delta_1 \Delta_2 . $$
When $ \Delta_1=0$, then the above inequality immediately implies   the bound (\ref{TEB}). Thus, without loss of generality, we assume that $ \Delta_1\not =0 .$
Denote by  $ r= \Delta_2/ \Delta_1.$ By substituting $ \Delta_2= r \Delta_1 $ into the above inequality, we have
\begin{equation} \label{III03}
\| x-{\cal H}_k(z)\|_2^2  \leq  (1+2r)\Delta_1^2 + C^2 . \end{equation}
We also note that  $ \Delta_1^2 + \Delta_2 ^2  \leq C^2 $ which together with  $ \Delta_2 = r \Delta_1$ implies that
$  \Delta _1 ^2 \leq  C^2/(1+ r^2). $
Thus it follows from (\ref{III03}) that
 \begin{equation} \label{III06}  \| x-{\cal H}_k(z)\|_2^2  \leq \left( 1+  \frac{1+2r}{1+r^2}   \right)  C^2 = g(r)  C^2,  \end{equation}
where $$  g(r): = 1+ \frac{1+2r}{1+r^2}= \frac{2(1+r)+r^2}{ 1+r^2} . $$
Consider the maximum of $g(r) $ over the interval $[0, \infty). $ If $r=0,$ then $ g(0) = 2. $ When $r \to \infty, $ we see that  $ g(r) \to 1. $ Note that $g(r)$ has a unique stationary point in  $ [0, \infty),$ i.e., the equation
$ 0=g'(r) = \frac{2(1-r-r^2)} {(1+r^2)^2}  $ has  a unique solution given by
 $r^*= \frac{\sqrt{5}-1}{2} \approx 0.618 $   at which
$$ g(r^*) = 1+  \frac{1+2r^*}{1+(r^*)^2} =\frac{5+\sqrt{5}}{5-\sqrt{5}} =  (\frac{\sqrt{5}+1}{2} )^2. $$ Thus the maximum value of $g(r)$ over the interval $[0, \infty)$ is given by $$ \max\{g(0), g(r^*), g(\infty)\}= g(r^*)=  (\frac{\sqrt{5}+1}{2} )^2 .$$  Therefore it follows from (\ref{III06}) that
$$ \|x-{\cal H}_k(z) \|_2 \leq  \sqrt{g(r^*)} C =  \frac{\sqrt{5}+1}{2} C, $$
which is the desired relation (\ref{TEB}).  \hfill $\Box $

\vskip 0.1in

\emph{Note:} After the first version of the manuscript appeared in arXiv, J. Shen communicated to us to point out that Lemma 2.2 above can actually follow from Shen and Li's Theorem 1 in \cite{SL18} which claims that for any vector $b \in \mathbb{R}^n $ and $ k$-sparse vector $x \in \mathbb{R}^n $ and for any $ q\geq k, $ one has
$$ \|{\cal H}_q (b)-x\|_2 \geq \sqrt{\mu} \|b-x\|_2,  ~ \mu = 1+\frac{\rho+\sqrt{(4+\rho)\rho}}{2}, $$ $$\rho = \frac{\min\{k, n-q\}} { q-k +  \min\{k, n-q\} }. $$
From such a result, it is not difficult to show that the bound in Lemma 2.2 can be also obtained from Theorem 1 in \cite{SL18}.

%\textbf{It is worth mentioning that the bound (\ref{TEB}) is tighter than the one established recently by Shen and Li \cite{SL18} who provided a bound for the term $\|x-{\cal H}_k(z) \|_2 $ %in terms of $ \| x- z\|_2 $ which, in general,  is looser than the term  $ \| (x- z)_{S\cup S^*}\|_2.  $    Later we will find that the   estimation (\ref{TEB})  plays a key role in %establishing the improved performance results for IHT and  CoSaMP algorithms.  We now  point out that the bound (\ref{TEB}) is the tightest of its kind (and hence it cannot be improved %further). It is this tightness that makes it possible to improve the RIP-based bounds for the guaranteed performance of signal recovery with the above-mentioned algorithms.}

\vskip 0.1in

\begin{Exam} \label{EXAM}  [Tightness of (\ref{TEB})].  Let  $ 0< \tau < k$ be two given integer numbers.   Consider two vectors in $ \mathbb{R}^n ~ (n> k+\tau) $ of the following form:
\begin{align*}  z  & =(\overbrace{1,   \dots, 1}^{k}, \overbrace{\varepsilon, \dots, \varepsilon}^\tau, 1/2, \dots, 1/2)^T  \in \mathbb{R}^n, \\
 x & =  (\overbrace{0, \dots, 0}^\tau, \overbrace{1, \dots, 1}^{k-\tau},  \overbrace{\alpha+\varepsilon, \dots, \alpha+ \varepsilon}^{\tau}, 0, \dots, 0)^T  \in \mathbb{R}^n ,
\end{align*}
where $ \alpha \geq 0$ and $ 0< \varepsilon \leq 1 $ are two parameters.
\end{Exam}

For the vectors $x$ and $z$ given above,
we see that $x$ is   $k$-sparse   and we may
take
$$ {\cal H}_k(z) = (\overbrace{1, \dots, 1}^k, \overbrace{0, \dots, 0}^{n-k} ) ^T \in \mathbb{R}^n.  $$
 Clearly, $ S^*= \text{supp} ({\cal H}_k(z)) = \{1, \dots, k \}$ and $ S=\text{supp} (x) = \{ \tau+1, \dots, \tau+ k\}.$ Thus
$$ S^* \cup S = \{1, 2, \dots, k+ \tau \}, ~ S^* \cap S = \{ \tau +1, \dots, k \}, $$
and hence
$$  \|(x-z)_{S^* \cup S} \|_2^2 = \tau (1+ \alpha^2), $$
$$ \|x-{\cal H}_k(z)\|_2^2 = \tau \left[1+ (\alpha+\varepsilon)^2\right]. $$
Consider the ratio
$$  \frac{\|x-{\cal H}_k(z)\|_2^2 } { \|(x-z)_{S^* \cup S} \|_2^2 } = \frac{1+ (\alpha+\varepsilon)^2} { 1+ \alpha^2} =:   g (\alpha, \varepsilon). $$
We now find the maximum value of the function $ g(\alpha, \varepsilon)$ with respect to $ \alpha \in [0, \infty). $ It is easy to check that there exists a unique stationary point of $g(\alpha, \varepsilon)$ with respect to $\alpha\in [0, \infty). $ In fact, let
$ \frac{\partial g(\alpha, \varepsilon)}{ \partial \alpha} =0$ which leads to
$ \alpha^2  +  \alpha \varepsilon  -1 =0.$
Thus the unique stationary point of $ g(\alpha, \varepsilon)$ in $[0, \infty) $ is
$ \alpha^*= \frac{\sqrt{4+\varepsilon^2}-\varepsilon}{2},  $ at  which
\begin{align*} g(\alpha^*, \varepsilon)  & =   \frac{1+(\alpha^*+\varepsilon)^2} {1+(\alpha^*)^2}
 =   \frac{1+ (\frac{\sqrt{4+\varepsilon^2}+\varepsilon}{2})^2} {1+(\frac{\sqrt{4+\varepsilon^2}-\varepsilon}{2} )^2}  =    \frac{ 1+ \sqrt{\frac{\varepsilon^2}{4+\varepsilon^2} }} {  1- \sqrt{\frac{\varepsilon^2}{4+\varepsilon^2} }}\\
 &= g_1(g_2(\varepsilon)),
\end{align*}
where the functions $ g_1 $  and $ g_2 $ are defined as follows:
$$ g_2(\varepsilon) =  \sqrt{\frac{\varepsilon^2}{4+\varepsilon^2} }, ~~ g_1(t) = \frac{ 1+t}{1-t}, $$
where $ 0\leq t< 1.$ Clearly, $g_1$ and $ g_2$ are increasing functions  and $g_2(\varepsilon ) < 1.  $   Thus $   g(\alpha^*, \varepsilon) $ is an increasing function of $\varepsilon$ over $(0, 1]. $ Therefore, as $ \varepsilon$   takes a  value  close to $1,$  the maximum of the function is achieved at $ \varepsilon=1. $  Note that
$ g(\alpha^*, \varepsilon)  = \frac{ \sqrt{4+\varepsilon^2}+\varepsilon }  {  \sqrt{4+\varepsilon^2}-\varepsilon }.$
 Thus
$$\lim_{\varepsilon \to 1} g(\alpha^*, \varepsilon) =\frac{ \sqrt{5}+1} {\sqrt{5}-1 } = ( \frac{\sqrt{5}+1}{2} )^2 \geq 1+\varepsilon^2  $$ for any $ \varepsilon \in (0,1].$
As $ g(0, \varepsilon)= 1+ \varepsilon^2$ and $ g(\infty, \varepsilon) := \lim_{\alpha \to \infty} g(\alpha, \varepsilon) =1, $
the  maximum of $ g(\alpha, \varepsilon) $ in $[0, \infty)$ is determined as follows:
\begin{align} \label {ggga}    \max_{\alpha\in [0, \infty)} g(\alpha, \varepsilon)
& = \max \{g(0, \varepsilon), g(\infty, \varepsilon), g( \alpha^* \varepsilon)\} \nonumber \\
& =   \max\{ 1+\varepsilon^2,  1 , g(\alpha^*, \varepsilon)\} \nonumber \\
&=  g(\alpha^*, \varepsilon),
\end{align}
 which tends to $( \frac{\sqrt{5}+1}{2} )^2   $ as $ \varepsilon \to 1.$
 This means the bound (\ref{TEB}) is tight since the ratio $ g(\alpha, \varepsilon) $ can approach  to $( \frac{\sqrt{5}+1}{2} )^2   $ for any level of accuracy provided that $ \alpha$ and $ \varepsilon$ are suitably chosen. In particular, this ratio can achieve the exact value $ ( \frac{\sqrt{5}+1}{2} )^2 $ by taking
$ \varepsilon =1$ and $ \alpha = \frac{\sqrt{5}-1}{2}. $  In other words,  the equality in (\ref{TEB}) can be  achieved at the vectors
$$ z = (\overbrace{1, \dots, 1}^{k+\tau}, 1/2, \dots, 1/2 )^T \in \mathbb{R}^n,  $$
$$ x= (\overbrace{0, \dots, 0}^{\tau}, \overbrace{1, \dots, 1}^{k-\tau}, \overbrace{\eta, \dots, \eta}^{\tau}, 0, \dots, 0)^T \in \mathbb{R}^n , $$
where $ \eta= (\sqrt{5}+1)/2.$  \hfill $\Box $

\vskip 0.1in

In the rest of the paper, we will use the following   concept of restricted isometry constant (RIC) and its several  useful properties listed in   Lemma \ref{Lem-Basic} below.

\vskip 0.1in

\begin{Def}  \cite{CT05}  Let $A $ be a given $m\times n$  matrix  with $m <n. $  The  restricted isometry constant (RIC),  denoted $\delta_q: = \delta_q (A), $ is the smallest number $\delta \geq 0$ such that
$$ (1-\delta) \|x\|^2_2 \leq \|Ax\|^2_2
\leq (1+\delta)\|x\|^2_2 $$ holds  for all $q$-sparse vectors
$x\in \mathbb{R}^n. $ If $ \delta_q<1,$ then $A$ is said to satisfy the restricted isometry property (RIP) of order $q. $
\end{Def}

From the definition,  we see that $ \delta_{q_1} \leq \delta_{q_2} $ for $ q_1\leq q_2.$
Implied directly from the above definition are the following properties which are widely utilized in the compressed sensing literature and in this paper.

\vskip 0.1in

\begin{Lem}  \label{Lem-Basic}  \cite{CT05, F11,NT09}   (i)  Let $u,v \in \mathbb{R}^n $ be $s$-sparse and $t$-sparse  vectors, respectively. If $  \textrm{supp}   (u) \cap  \textrm{supp}  (v) = \emptyset, $ then $$|u^T A^T A v | \leq \delta_{s+t} \|u\|_2 \|v\|_2. $$

 (ii) Let $  v \in \mathbb{R}^n $ be a vector and   $S \subseteq \{1,2, \dots, n\}$ be an index set. If $|S\cup  \textrm{supp}  (v) | \leq t,$ then
$$ \| [(I-A^TA)v]_S\|_2  \leq \delta_t \|v\|_2 .
$$

(iii) Let $ \Lambda \subseteq \{1, \dots, n\}$ be an index set satisfying that $ \Lambda \cap \textrm{supp} (u) = \emptyset$ and  $ | \Lambda \cup  \textrm{supp} (u) | \leq t.$ Then $$ \|(A^T A u)_\Lambda\|_2\leq \delta_t \| u\|_2.    $$
\end{Lem}

Item (iii) follows from (ii). In fact,   when $\Lambda \cap  \textrm{supp}  (u) = \emptyset$ which means $ u_{\Lambda}=0,$  one has $ \|(A^T A u)_\Lambda\|_2= \|[(I-A^TA)u]_\Lambda \|_2 \leq \delta_t \|u\|_2.$

We now start to establish an  improved result for the guaranteed performance of signal recovery via IHT. The result for CoSaMP will be given separately in section \ref{section3}. Such improvements in terms of RIP are vital for both compressed sensing theory and algorithms. The RIP-based bound directly clarifies the scenarios in which the algorithms are guaranteed to be successful in signal recovery.  A more relaxed RIP condition is imposed, the broader the class of signal recovery problems that can be solved successfully by the algorithms are identified. Moreover, the relaxed RIP-based bound   can also dramatically impact  on the number of measurements required for signal recovery. As shown in \cite{ CT06, MPT08, BDDW08},  for Gaussian random sensing matrix $A$ of size $ m \times n ~ (m \ll n)$, there is a universal constant $ C^* > 0$ such that the RIC of $A/\sqrt{m} $ satisﬁes $\delta_{2k} \leq \delta^*  < 1 $  with probability at least $ 1-\xi $ provided that $$ m\geq C^* (\delta^*)^{-2} (k (1+\ln (n/k))+\ln (2\xi -1)).$$ From this result, it can be seen that the higher the  bound $\delta^* ,$ the less number of measurements is required.

It is worth mentioning that the practical signal $ x$ may not necessarily be $k$-sparse. Let  $S$ be the index set for the largest $k$ absolute entries  of the signal $ x,$ i.e.,  $ S=L_k(x). $ Then the $k$-sparse vector $x_S$ is  the best $k$-term approximation of $x.$ In terms of $x_S,$ $y=Ax+\nu = Ax_S + \nu' $ with $ \nu'= Ax_{\overline{S}} + \nu . $  This means the measurements $y$ of the original signal $x$ can be seen as the measurements of the $k$-sparse signal $x_S$ with measurement error $\nu'.$ Thus when the signal is not $k$-sparse, the recovery can be made for  a smaller number of  significant components of the target signal, i.e., only the $k$ terms of the signal are recovered.

We are ready to show the  main result for IHT.

\vskip 0.1in

\begin{Thm} \label{Thm-IHT} Suppose that the sensing matrix $A$ satisfies
$$ \delta_{3k} < \frac{\sqrt{5}-1}{2} \approx 0.618.$$ Let $ y= Ax+\nu $ be the measurements of $x$  with error $ \nu$ and $S=L_k (x).$  Then the iterates $x^p,$ generated by the IHT,   approximate $x$ with error
\begin{equation} \label{ERR-IHT} \|x^p-x_S\|_2 \leq \rho^p \| x^0- x_S\|_2+  \frac{\sqrt{5}+1}{2 (1-\rho)} \| A^T  \nu '\|_2 ,\end{equation}  where $ \nu'= A x_{\overline{S}}+\nu, $ and the constants $ \rho $  is given as
\begin{equation} \label{ccrrr} \rho = \frac{\sqrt{5}+1}{2}\delta_{3k} <1.
\end{equation}

\end{Thm}

{\it Proof.}  Denote by $ u^p:=x^p+ A^T (y- A x^p). $   By the structure of the IHT, $  S^{p+1} := \textrm{supp} (x^{p+1})= \textrm{supp} ({\cal H}_k(u^p)). $  By Lemma 2.2, one has
\begin{align} \label{EE*} \| x_S- x^{p+1} \|_2  &  =\| x_S- {\cal H}_k(u^p) \|_2   \nonumber \\
 &  \leq \frac{\sqrt{5}+1}{2} \| (x_S- u^p)_{S^{p+1}\cup S}\|_2. \end{align}
We now estimate the term $ \| (x_S- u^p)_{S^{p+1}\cup S}\|_2$ which can be bounded as follows:
\begin{align*} &  \| (x_S- u^p)_{S^{p+1}\cup S}\|_2 \\ & =    \| (x_S-x^p-A^T (y- A x^p))_{S^{p+1}\cup S}\|_2\\
 & =
 \| (x_S-x^p-A^T (Ax_S + \nu'- A x^p))_{S^{p+1}\cup S}\|_2\\
& =
\| [(I-A^T A) (x_S-x^p)-A^T \nu' ]_{S^{p+1}\cup S}\|_2\\
& \leq  \| [(I-A^T A) (x_S-x^p)]_{S^{p+1}\cup S}\|_2 + \|A^T \nu'\|_2\\
& \leq   \delta_{3k} \|x_S-x^p\|_2 + \|A^T \nu'\|_2,
\end{align*}
where the last inequality follows from Lemma \ref{Lem-Basic} (ii) with $| \textrm{supp} (x_S-x^p) \cup (S^{p+1} \cup S)| \leq 3k. $
 Substituting this into (\ref{EE*}) yields
\begin{align}  \label{RHO}    \| x_S     - x^{p+1} \|_2
&  \leq \frac{\sqrt{5}+1}{2} (\delta_{3k} \|x_S-x^p\|_2 +    \|A^T \nu'\|_2) \nonumber\\
 & = \rho \|x_S-x^p\|_2  +   \frac{\sqrt{5}+1}{2}\|A^T \nu'\|_2,
 \end{align}
 where  the constant $$ \rho = \frac{\sqrt{5}+1}{2}\delta_{3k} <1, $$ provided that $$ \delta_{3k} < \frac{2} {\sqrt{5}+1}= \frac{\sqrt{5}-1}{2} \approx 0.618.$$
The error bound (\ref{ERR-IHT}) immediately follows from (\ref{RHO}). \hfill $\Box $

\vskip 0.1in

\begin{Rem}  The above result improves the current best known bound  $ \delta_{3k} < \frac{1}{\sqrt{3}}\approx 0.5774 $ for IHT established by Foucart and Rauhut \cite{FR13, F11}. The tightness of the relation (\ref{TEB}) (as indicated by Example \ref{EXAM}) is essential to the  improvement of the RIP-based bound for IHT. The tightness of  (\ref{TEB}) and the simple argument in the proof of Theorem \ref{Thm-IHT}  point to the conjecture that the new bound $\delta_{3k} < (\sqrt{5}-1)/2$ for IHT is optimal. This, however, is an interesting conjecture requiring a further investigation.   It is also worth mentioning that some researchers developed the RIP-based bounds for the performance of compressed sensing algorithms according to the geometric rate $\rho\leq 0.5$ instead of $ \rho <1. $ From the analysis above, if we require the geometric rate $\rho$ given in (\ref{ccrrr}) be less than 0.5, namely,
$\rho = \frac{\sqrt{5}+1}{2}\delta_{3k}  \leq 0.5,
$
 which is guaranteed by $ \delta_{3k} \leq (\sqrt{5}-1)/4 \approx 0.309,$  then we immediately obtain from (\ref{RHO}) the following recovery error:
$$\|x^p-x_S\|_2 \leq 0.5^p \| x^0- x_S\|_2+ (\sqrt{5}+1)\| A^T  \nu '\|_2. $$ We clearly see that our result for IHT also remarkably improves the existing result $ \delta_{3k} \leq 0.22 $ established  by Shen and Li \cite{SL18} for  IHT   in terms of geometric rate $0.5 .$   From the proof of Theorem 6.18 in \cite{FR13}, it is easy to verify that the RIP-based bound obtained by Foucart and Rauhut is $ \delta_{3k} \leq \frac{1}{2\sqrt{3}}\approx 0.2886 $ in terms of geometric rate  $0.5.$

\vskip 0.1in

The estimation (\ref{ERR-IHT}) implies the finite termination and stability of IHT through a standard lemma such as Lemma 6.23  in \cite{FR13}.  We don't state the stability results and the ones concerning the number of iterations required to achieve the desired recovery accuracy (the interested reader can see the statement of such results in \cite{FR13, BFH16} for details). In   this paper, we only focus on the establishment of the estimation like (\ref{ERR-IHT}) which ensures the convergence of an algorithm and the success of signal recovery/approximation with the algorithm.

\end{Rem}

\section{Improved RIP bound for CoSaMP}\label{section3}

As pointed out in \cite{Z20, ZL20}, the hard thresholding operator may cause numerical oscillation, and thus the  IHT may fail to consistently reduce the objective value of the model (\ref{L0}) during the course of iterations. The orthogonal projection is one of the techniques which may alleviate the oscillation problem. Thus it is widely used in hard-thresholding-based algorithms   including CoSaMP, SP, the latest Newton-step-based thresholding \cite{MZ20}, and optimal $k$-thresholding algorithms \cite{Z20, ZL20} .
In this section, we show the main result for CoSaMP. Before doing do, we first state a few technical results.

\vskip 0.1in

\begin{Lem}\label{Lem-3A}
Given three constants $ \alpha_1, \alpha_2, \alpha_3\geq 0$ where $ \alpha_1<1,$ if $t$ satisfies the  condition
$ 0\leq t-\alpha_3\leq \alpha_1 \sqrt{t^2+\alpha_2^2},$ then
$$ t\leq \frac{\alpha_1}{\sqrt{1-\alpha_1^2}} \alpha_2 + \frac{1}{1-\alpha_1}\alpha_3. $$
\end{Lem}

\emph{Proof}.  Under the conditions of the Lemma,  $t$ satisfies the condition
$ (t-\alpha_3)^2 \leq \alpha_1^2 (t^2+\alpha_2^2) $, i.e.,
$$ \phi(t) := (1-\alpha_1^2) t^2 -2t \alpha_3 +\alpha_3^2-\alpha_1^2 \alpha_2^2 \leq 0.$$
Thus $t$ is less than or equal to the largest real root of the quadratic equation $\phi (t)=0. $ That is,
 \begin{align*}
t & \leq  \frac{2\alpha_3+ \sqrt{4 \alpha_3^2-4 (1-\alpha_1^2)(\alpha_3^2-\alpha_1^2 \alpha_2^2)}}{ 2 (1-\alpha_1^2)} \\
 & =    \frac{\alpha_3+ \sqrt{ \alpha_1^2\alpha_3^2+ (1-\alpha_1^2)\alpha_1^2 \alpha_2^2}}{   1-\alpha_1^2 } \\
& \leq   \frac{\alpha_3+  \alpha_1\alpha_3+ \alpha_1 \alpha_2\sqrt{1-\alpha_1^2}} {1-\alpha_1^2}  \\
& =  \frac{\alpha_1}{\sqrt{1-\alpha_1^2} } \alpha_2+\frac{\alpha_3}{1-\alpha_1},
\end{align*}
as desired. \hfill $\Box $

A fundamental property of the orthogonal projection is given as follows. A similar property can be found in the literature, however, the following one is more general than the existing ones.

\vskip 0.1in

\begin{Lem} \label{Lem-3B} Let $y=Ax+\nu$ be  the measurements of the signal $x$ where $\nu $ is a noisy vector. Let $S, \Lambda \subseteq\{1, \dots,n\}$ be two nonempty index sets and $|S|\leq \tau $ where $ \tau $ is an integer number.  Let $x^*$ be the solution to the orthogonal projection problem
\begin{equation} \label{OP} x^*=\arg\min_{z\in \mathbb{R}^n} \{\|y-A z\|_2: ~   \textrm{supp}   (z) \subseteq  \Lambda\}. \end{equation}   Let $\Gamma$ be any given index set satisfying $ \Lambda \subseteq  \Gamma \subseteq \{i:   \left[A^T ( y- A x^*)\right]_i =0\}. $ If $\delta_{|\Gamma| +\tau } <1, $ then
\begin{equation} \label{B01} \| ( x_S-x^*)_{\Gamma} \|_2  \leq   \frac{ \delta_{|\Gamma|+\tau}   \|( x_S-x^*)_{\overline{\Gamma}}\|_2 }{\sqrt{1-\delta^2_{|\Gamma| +\tau }}}   + \frac{\|A^T \nu'\|_2}{1-\delta_{|\Gamma|+\tau}}, \end{equation}
and hence
\begin{equation} \label{B02} \|x_S-x^*\|_2 \leq \frac{\|( x_S-x^*)_{\overline{\Gamma}}\|_2}{\sqrt{1-\delta^2_{|\Gamma| +\tau  }}}    + \frac{\|A^T \nu'\|_2}{1-\delta_{|\Gamma|+\tau}}, \end{equation}
where $ \nu'= Ax_{\overline{S}}+ \nu.$

\end{Lem}

\vskip 0.1in

{\it Proof. } Since $x^*$ is the optimal solution to the problem (\ref{OP}), by optimality, we immediately see that $ [A^T (y-Ax^*)]_\Lambda =0$ and thus the set $  \{i:   \left[A^T ( y- A x^*)\right]_i =0\}$ is nonempty since it contains $ \Lambda$ as a subset.   By the definition of $ \Gamma$, we have
$ \left[
A^T ( y- A x^*)\right]_{\Gamma}=0 $ which, together with $ y= A x_S + \nu' $  where $ \nu' = A x_{\overline{S}}+ \nu, $ implies that
\begin{align} \label{DDFF}
0 & = \left[A^T A(x_S- x^*)+ A^T \nu' \right]_{\Gamma}  \nonumber \\
 & =  \left[(A^T A-I)(x_S- x^*)\right]_{\Gamma} +
 (x_S-x^*)_{\Gamma} \nonumber \\
 & ~~~ + [A^T \nu'] _{\Gamma}.
\end{align}
As $ {\rm supp}(x_S-x^*) \subseteq S\cup \Lambda $ and $ \Lambda \subseteq \Gamma, $ we see that $$ |{\rm supp} (x_S-x^*) \cup \Gamma | \leq | (S\cup \Lambda)\cup \Gamma |= |S \cup \Gamma|\leq |\Gamma|+\tau. $$ Thus it follows from (\ref{DDFF}) and Lemma \ref{Lem-Basic} (ii) that
\begin{align}  \label{QQQY}  \| ( x_S    -      x^*)_{\Gamma} \|_2
 & \leq  \| \left[(A^T A-I)(x_S- x^*)\right]_{\Gamma}\|_2+ \|A^T \nu'\|_2  \nonumber \\
 & \leq \delta_{|\Gamma|+\tau} \| x_S-x^*\|_2 + \|A^T \nu'\|_2  \nonumber \\
 & = \delta_{|\Gamma|+\tau} \sqrt{  \| ( x_S-x^*)_{\Gamma} \|_2^2 +    \| ( x_S-x^*)_{\overline{\Gamma}} \|_2^2}  \nonumber \\
 & ~~~  + \|A^T \nu'\|_2.
\end{align}
If $ \|(x_S-x^*)_\Gamma \|_2\leq  \| A^T \nu'\|_2, $  the desired relations (\ref{B01}) and (\ref{B02}) hold trivially. Otherwise if $\| (x_S-x^*) _{\Gamma}\|_2 > \|A^T \nu\|_2,$
then by setting $ \alpha_1 = \delta_{|\Gamma|+\tau} <1, \alpha_2= \|(x_S-x^*)_{\overline{\Gamma}}\|_2, \alpha_3 =\|A^T \nu' \|_2, $ and $ t= \| (x_S-x^*)_{\Gamma}\|_2, $
it follows from (\ref{QQQY}) and  Lemma \ref{Lem-3A} that
\begin{equation} \label{ERROR-OP}  \| ( x_S-x^*)_{\Gamma} \|_2  \leq   \frac{\delta_{|\Gamma|+\tau}  \|( x_S-x^*)_{\overline{\Gamma}}\|_2 }{\sqrt{1- \delta^2_{|\Gamma| +\tau } }}   + \frac{\|A^T \nu'\|_2}{1-\delta_{|\Gamma|+\tau}}. \end{equation}
Note that \begin{equation} \label{abc}  \sqrt{(a+b)^2+ c^2} \leq \sqrt{a^2+c^2} + b     \end{equation}   for any  $ a,b,c \geq 0. $ It follows from (\ref{ERROR-OP}) and (\ref{abc}) that
\begin{align*}
& \|x_S-x^* \|_2^2 \\  & =    \|(x_S-x^*)_{\Gamma} \|_2^2 + \| (x_S- x^*)_{\overline{\Gamma}}\|_2^2  \\
 & \leq    \Big{(} \frac{\delta_{|\Gamma|+\tau} \|( x_S-x^*)_{\overline{\Gamma}}\|_2}{\sqrt{1-\delta^2_{|\Gamma| +\tau } }}  + \frac{\|A^T \nu'\|_2}{1-\delta_{|\Gamma|+k}} \Big{)} ^2  + \| (x_S- x^*)_{\overline{\Gamma}}\|_2^2  \\
&  \leq  \Big{ ( }     \sqrt{ \frac{\delta^2_{|\Gamma| +\tau }\|( x_S-x^*)_{\overline{\Gamma}}\|_2^2}{1-\delta^2_{|\Gamma| +\tau }}  +\| (x_S- x^*)_{\overline{\Gamma}}\|_2^2}  \\
&  ~~~~~ + \frac{\|A^T \nu'\|_2}{1-\delta_{|\Gamma|+\tau}}        \Big{ )  }^2 \\
 & =  \Big{( } \frac{1}{\sqrt{1-\delta^2_{|\Gamma| +\tau } }}  \|( x_S-x^*)_{\overline{\Gamma}}\|_2 + \frac{\|A^T \nu'\|_2}{1-\delta_{|\Gamma|+\tau}} \Big{)} ^2,
\end{align*}
as desired. \hfill $\Box $

 The next helpful technical result which together with Lemma \ref{Lem-3B}   eventually yields an improved RIP-based bound for the guaranteed success of  CoSaMP.

\vskip 0.1in

\begin{Lem} \label{Lem-Key}
Let $ y= Ax+\nu $ be the measurements of $x$ with measurement error $ \nu .$  Let $S= L_k (x).$  Given a $k$-sparse vector $ x^p$ with $ S^p= {\rm supp} (x^p)$ and the index set $$T=L_\beta(A^T(y-Ax^p)), $$ where  $ \beta \geq 2k $ is an integer number, if $ \delta_{2k+\beta}<1 $  then one has
\begin{align}\label{EELL}    \| (x^p  &  -x_S )_{\overline{T}}\|_2   \leq \sqrt{2} (\delta_{2k+\beta} \| x^p-x_S\|_2   +   \| A^T \nu'\|_2) ,\end{align}
where $ \nu' = A x_{\overline{S}}+ \nu .$
\end{Lem}

{\it Proof. } Let $S, S^p, $ and $T$ be defined as in the lemma.  If $ S \cup S^p \subseteq T, $ then $ \overline{T} \subseteq \overline{S\cup S^p}$ which implies that  $ \| (x^p -x_S )_{\overline{T}}\|_2  \leq  \| (x^p -x_S )_{\overline{S\cup S^p}}\|_2 =0. $ Thus the relation (\ref{EELL}) holds trivially. We  only need to show  (\ref{EELL}) for the case $ S\cup S^p \not\subseteq T . $  Thus in the remaining proof, we assume that $S\cup S^p \not\subseteq T.$  It is convenient to define
\begin{equation}    \label{EEHH01}  \Omega:= \|[ A^T (y-Ax^p)]_{(S\cup S^p)\backslash T} \|_2. \end{equation}
As the cardinality $|T| =\beta \geq 2k\geq |S\cup S^p|,$ we see that
\begin{align} \label{DDD}  |(S\cup S^p)\backslash T| & = |S\cup S^p|- |(S\cup S^p) \cap T| \nonumber \\
  & \leq |T|-|(S\cup S^p) \cap T| \nonumber \\
   & =|T\backslash (S\cup S^p)|. \end{align}
This means the number of the indices in $ (S\cup S^p )\backslash T$ is less than or equal to the number of elements in $T\backslash (S\cup S^p).$
By the definition of $T,$ the entries of the vector $A^T (y-Ax^p)$ supported on $(S\cup S^p)\backslash T$ are not among the $\beta$ largest absolute entries of the vector. This  together with (\ref{DDD})  implies that
 \begin{align}\label{EEHH02}  \|[A^T (y       -A     x^p)]_{(S\cup S^p)\backslash T}\|_2
  \leq  \|[A^T (y-Ax^p)]_{T \backslash (S\cup S^p)}\|_2. \end{align}
Denote by
$$ \Omega^* = \|[(x_S-x^p) -A^T (y- Ax^p)]_{(S\cup S^p) \Delta T}\|_2, $$
where $ \Delta$ denotes the symmetric difference of two sets, i.e.,
$$ (S\cup S^p) \Delta T = ((S\cup S^p)\backslash  T) \cup ( T\backslash (S\cup S^p)). $$
Note that
$$ |(S\cup S^p) \cup ((S\cup S^p) \Delta T)|\leq  |(S\cup S^p) \cup T|\leq 2k+\beta.$$ As $ y= A x_S+ \nu',$ by  Lemma \ref{Lem-Basic} (iii), we have
\begin{align} \label{OM*}   \Omega^*  & = \|[(I -A^T A)(x_S- x^p) + A^T \nu']_{(S\cup S^p) \Delta T} \|_2 \nonumber \\
& = \|[(I-A^T A) (x_S-x^p)]_{(S\cup S^p) \Delta T}\|_2  \nonumber \\
 & ~~~~  + \|(A^T \nu')_{(S\cup S^p) \Delta T}\|_2  \nonumber \\
 & \leq \delta_{2k+\beta} \|x_S-x^p\|_2 + \|A^T \nu'\|_2.
 \end{align}
Let $ \widehat{\Omega } = \|[A^T (y- Ax^p)]_{T\backslash (S\cup S^p)}\|_2 $
 and
 $$ W= (x_S-x^p)_{  (S\cup S^p)\backslash T } - [ A^T (y-Ax^p)]_{(S\cup S^p)\backslash T} . $$
 Then by the definition of $ \Omega^*,$ we see that
 \begin{align} \label{OM2} ( \Omega^*) ^2  & =  \|[(x_S-x^p)-A^T (y- Ax^p)]_{(S\cup S^p) \backslash T}\|_2 ^2 \nonumber  \\
  & ~~~~ +\|[(x_S-x^p) -A^T (y- Ax^p)]_{T\backslash (S\cup S^p)}\|_2^2
 \nonumber \\
 & =   \|[(x_S-x^p)-A^T (y- Ax^p)]_{(S\cup S^p) \backslash T}\|_2 ^2 \nonumber  \\
   &  ~~~ +\|[A^T (y- Ax^p)]_{T\backslash (S\cup S^p)}\|_2^2 \nonumber \\
 & = \|W\|_2^2 + \widehat{\Omega}^2
 \end{align}
 where the  second equality follows from the fact $(x_S-x^p)_{T\backslash (S\cup S^p)}=0. $ There are only two cases.

 Case 1:  $\widehat{\Omega} =0. $ Then by (\ref{EEHH02}), we must have $ [ A^T (y-Ax^p)]_{(S\cup S^p)\backslash T} =0. $   From the definition of $W$ and the relation (\ref{OM*}) and noting that $ \|(x_S-x^p)_{\overline{T}}\|_2 =  \|(x_S-x^p)_{(S\cup S^p)\backslash T}\|_2, $  we immediately have that
$$  \|(x_S-x^p)_{ \overline {T}}\|_2 = \|W\|_2 \leq \Omega^* \leq   \delta_{2k+\beta} \|x_S-x^p\|_2 + \|A^T \nu'\|_2.$$ Thus the bound (\ref{EELL}) holds trivially for this case.

Case 2:  $ \widehat{\Omega} \not =0. $ Then let $ \beta $ be  the ratio of  $ \|W\|_2$  and $ \widehat{\Omega}, $ i.e.
 $\|W\|_2= \beta \widehat{\Omega}.  $ Substituting this into (\ref{OM2}), we immediately obtain
 \begin{equation} \label{WWEE}  \widehat{\Omega} = \frac{1}{\sqrt{1+\beta^2}} \Omega^*, ~~ \|W\|_2 =  \frac{\beta}{\sqrt{1+\beta^2}} \Omega^*. \end{equation}
 Therefore,
 \begin{align}  \label{2828}   \Omega ^2        & = \| (x_S-x^p)_{(S\cup S^p) \backslash T}   \nonumber  \\
 &   ~~~   -  \left[(x_S-x^p)_{ (S\cup S^p) \backslash T } - [ A^T (y-Ax^p)]_{(S\cup S^p) \backslash T}\right] \|_2^2 \nonumber \\
& = \| (x_S-x^p)_{(S\cup S^p) \backslash T} - W \|_2^2 \nonumber \\
& = \| (x_S-x^p)_{(S\cup S^p) \backslash T} \|_2^2-  2\left[(x_S-x^p)_{(S\cup S^p) \backslash T}\right]^T W \nonumber \\
& ~~~ + \|W\|_2^2 \nonumber  \\
& \geq \| (x_S-x^p)_{(S\cup S^p) \backslash T} \|_2^2 -  2 \| (x_S-x^p)_{(S\cup S^p) \backslash T}\|_2 \|W\|_2 \nonumber\\
 & ~~~ + \|W\|_2^2
\end{align}
By (\ref{EEHH02}) and (\ref{WWEE}), the  left-hand side of the above inequality can be bounded as $$ \Omega^2 \leq \widehat{\Omega} ^2= \frac{(\Omega^*)^2}{1+\beta^2}.$$   Thus substituting $ \|W\|_2$   in (\ref{WWEE}) into   (\ref{2828})  leads to
\begin{align} \frac{(\Omega^*)^2}{1+\beta^2}  & \geq \| (x_S-x^p)_{(S\cup S^p) \backslash T} \|_2^2 + \frac{\beta^2}{1+\beta^2}(\Omega^*)^2 \nonumber\\
& ~~~  -  2 \| (x_S-x^p)_{(S\cup S^p) \backslash T}\|_2 \frac{\beta}{\sqrt{1+\beta^2}} \Omega^*. \end{align}   Simplifying yields
\begin{align*} &  \| (x_S-x^p)_{(S\cup S^p) \backslash T} \|_2^2-   \frac{2\beta\Omega^* }{\sqrt{1+\beta^2}} \| (x_S-x^p)_{(S\cup S^p) \backslash T}\|_2   \\
 & ~~~~~ + \frac{(\beta^2-1)(\Omega^*)^2}{1+\beta^2} \leq 0,
 \end{align*}
which is a quadratic inequality of $ \| (x_S-x^p)_{(S\cup S^p) \backslash T} \|_2$, and thus
\begin{align*}  \| (x_S   &   -x^p)_{(S\cup S^p) \backslash T} \|_2  \nonumber \\
 & \leq \frac{ \frac{2\beta\Omega^* }{\sqrt{1+\beta^2}}   +  \sqrt{  \frac{4\beta^2(\Omega^* )^2}{1+\beta^2}  -\frac{4(\beta^2-1)(\Omega^*)^2}{1+\beta^2} }  } { 2}  = \frac{1+ \beta}{\sqrt{ 1+ \beta^2} } \Omega^* \nonumber \\
 & \leq  \left(\max_{0<\beta < \infty} \frac{1+ \beta}{\sqrt{ 1+ \beta^2} }
 \right)\Omega^* \nonumber \\
 & = \sqrt{2} \Omega^*,
 \end{align*}
where the maximum of the univariate function of $ \beta$ achieves at $ \beta =1. $ Combined with (\ref{OM*}), we immediately obtain the desired inequality in the lemma.  \hfill $\Box $

\vskip 0.1in

The main result for CoSaMP is summarized as follows.

\vskip 0.1in

\begin{Thm} \label{Thm-CoSaMP}
If the restricted isometry constant of the sensing matrix $A$ satisfies that
\begin{equation} \label{DDDLLL} \delta_{4k}< \sqrt{\frac{2}{\sqrt{13+4\sqrt{5}}+3}} \approx 0.5102, \end{equation}  then the iterates $ \{ x^p\}, $ generated by the CoSaMP, satisfy that
$$ \|x_S-x^p\|_2\leq \rho^p \|x_S-x^0\|_2 +  \frac{C}{1-\rho} \|A^T\nu'\|_2,  $$
 where the constants $ \rho $ and $ C$ are given as  $$ \rho =  \delta_{4k}   \sqrt{  \frac{ 2+ (\sqrt{5}+1) \delta_{4k}^2}{1-\delta^2_{4k }}  } <1 $$ and $$  C= \sqrt{  \frac{ 2+ (\sqrt{5}+1) \delta_{4k}^2}{1-\delta^2_{4k }}  }  +  \frac{\sqrt{5}+1}{2(1-\delta_{4k})} . $$
\end{Thm}

\emph {Proof.}  Let $ U^{p+1}, z^{p+1}$ and $ x^{p+1} $ are given, respectively, by the steps (CP1)--(CP3) of CoSaMP (Algorithm 2 in section \ref{section1}). From the structure of CoSaMP, we see that
$ S^p= \textrm{supp} (x^p) \subseteq U^{p+1}$ and $ S^{p+1}= \textrm{supp} (x^{p+1})= \textrm{supp} ({\cal H}_k( z^{p+1} )) \subseteq U^{p+1}$ (so $(x^{p+1})_{U^{p+1}} = x^{p+1}). $ By Lemma \ref{HHB}, we have
\begin{align*}
  \| (x_S  &   - x^{p+1})_{U^{p+1}}\|_2   =    \|x_{S\cap U^{p+1}}- x^{p+1} \|_2 \\
 & =   \| x_{S\cap U^{p+1}}- {\cal H}_k( z^{p+1} ) \|_2 \\
 & \leq   \frac{\sqrt{5}+1}{2} \| (x_{S\cap U^{p+1}}- z^{p+1})_{(S\cap U^{p+1})\cup S^{p+1}} \|_2  \\
 & \leq   \frac{\sqrt{5}+1}{2} \| (x_{S\cap U^{p+1}}- z^{p+1})_{U^{p+1} } \|_2 \\
 &  ~~~~~~~~~~~ \textrm{(since  }  (S\cap U^{p+1}) \cup S^{p+1} \subseteq  U^{p+1})\\
 & =   \eta \| (x_{S}- z^{p+1})_{U^{p+1}} \|_2 ,
\end{align*}
where $ \eta=(\sqrt{5}+1)/2.$
Also, since $ \textrm {supp} (x^{p+1}) \subseteq U^{p+1},$  we have  that $ ( x^{p+1})_{\overline{U^{p+1}}}=0=(z^{p+1})_{\overline{ U^{p+1}}} .$ This together with the above relation implies that
\begin{align} \label{DDDY}  & \| x_S     -x^{p+1}\|_2^2  \nonumber \\ & = \|(x_S- x^{p+1})_{\overline{U^{p+1}}}\|_2^2+ \| (x_S-x^{p+1})_{U^{p+1}}\|_2^2 \nonumber \\
 & = \|(x_S- z^{p+1})_{\overline{U^{p+1}}}\|_2^2+ \| (x_S-x^{p+1})_{U^{p+1}}\|_2^2  \nonumber \\
 & \leq  \|(x_S- z^{p+1})_{\overline{U^{p+1}}}\|_2^2  + [\eta \| (x_S-z^{p+1})_{U^{p+1}}\|_2]^2.
\end{align}
Consider the step (CP2) of the CoSaMP.
 Applying Lemma \ref{Lem-3B} with $\Gamma = \Lambda = U^{p+1}$, $ S= L_k(x), $  $ x^*= z^{p+1} , $  $ \tau =k $ and $ |\Gamma | =| U^{p+1}| \leq 3k, $ we conclude that if $ \delta_{4k} <1,$ one has
\begin{align} \label{DDDZ} \|(x_S  &   -z^{p+1})_{U^{p+1}}\|_2  \nonumber \\
 & \leq \frac{\delta_{4k} \|( x_S-z^{p+1})_{\overline{U^{p+1}}}\|_2 }{\sqrt{1-\delta^2_{4k }}}  + \frac{\|A^T \nu'\|_2}{1-\delta_{4k}}.
\end{align}
  By combining the two  relations (\ref{DDDY}) and (\ref{DDDZ}) and using the inequality (\ref{abc}), we obtain
\begin{align}   \label{HHOOWW} &  \| x_S-x^{p+1}\|_2^2
  \leq  \|(x_S- z^{p+1})_{\overline{U^{p+1}}}\|_2^2  \nonumber \\
  & ~~~~~~~~~~~~~ +\left(  \frac{\eta \delta_{4k}  \|( x_S-z^{p+1})_{\overline{U^{p+1}}}\|_2  }{\sqrt{1-\delta^2_{4k }}}
  + \frac{\eta\|A^T \nu'\|_2}{1-\delta_{4k}} \right) ^2 \nonumber \\
& \leq \left(   \sqrt{ ( 1  +     \frac{\eta^2 \delta_{4k}^2}{1-\delta^2_{4k }} )  \|( x_S-z^{p+1})_{\overline{U^{p+1}}}\|_2 ^2      }   +  \frac{\eta \|A^T \nu'\|_2}{1-\delta_{4k}}  \right) ^2   \nonumber\\
& =  \Big{(}   \sqrt{  \frac{ 1+   \eta \delta_{4k}^2}{1-\delta^2_{4k }}  } \|( x_S-z^{p+1})_{\overline{U^{p+1}}}\|_2    +  \frac{\eta\|A^T \nu'\|_2}{1-\delta_{4k}}\Big{)}^2,
\end{align}
where the equality follows from the fact $ \eta^2-1= \eta. $   In the remainder of this proof,  it is sufficient to bound the term  $ \|( x_S-z^{p+1})_{\overline{U^{p+1}}}\|_2 $ in terms of $ \| x_S-x^p\|_2.$
Note that  $ (z^{p+1})_{\overline{U^{p+1}}}=0$ and $ S^p= \textrm{supp} (x^p) \subseteq U^{p+1} $ which implies that $ (x^p) _ {\overline{U^{p+1}}}=0.$ Thus
\begin{align} \label{R11}  \|( x_S-z^{p+1})_{\overline{U^{p+1}}}\|_2   & = \| (x_S)_{\overline{U^{p+1}}}\|_2 =  \|( x_S-x^{p})_{\overline{U^{p+1}}}\|_2 . \end{align}
Setting $ \beta =2k$ and $ T= {\rm supp} [{\cal H}_\beta(A^T(y-Ax^p))],$ it follows from Lemma \ref{Lem-Key} that the CoSaMP satisfies the relation
\begin{equation}\label{R22} \| (x_S-x^p)_{\overline{T}}\|_2 \leq  \sqrt{2} (\delta_{4k}  \| x^p-x_S\|_2 +   \| A^T \nu'\|_2) . \end{equation}
Note that $ T \subseteq U^{p+1}$ which implies that $ \overline{U^{p+1}} \subseteq  \overline{T} . $ Thus
\begin{equation} \label{R33}  \|( x_S-x^{p})_{\overline{U^{p+1}}}\|_2 \leq \| (x_S-x^p)_{\overline{T}}\|_2. \end{equation}
Merging the above three relations (\ref{R11})--(\ref{R33}) leads to
$$  \|( x_S    -z^{p+1})_{\overline{U^{p+1}}}\|_2   \leq   \sqrt{2} ( \delta_{4k}  \| x^p-x_S\|_2 +  \| A^T \nu'\|_2 ).
$$
Therefore, it follows from (\ref{HHOOWW}) that
 \begin{align*}    \| x_S  -x^{p+1}\|_2
  & \leq    \sqrt{  \frac{ 2(1+ \eta \delta_{4k}^2)}{1-\delta^2_{4k }}  }  ( \delta_{4k}  \| x_S-x^p\|_2
    \\
    & ~~~~  +   \| A^T \nu' \|_2 )     +  \frac{ \eta}{1-\delta_{4k}}\|A^T \nu'\|_2  \\
& = \rho \| x_S-x^p\|_2  + C \|A^T \nu'\|_2,
\end{align*}
where the constants  $ \rho $ and $ C$ are given by
  $$ \rho      = \delta_{4k}   \sqrt{  \frac{ 2+ (\sqrt{5}+1) \delta_{4k}^2}{1-\delta^2_{4k }}  }.  $$
  \begin{align*} C   & =     \sqrt{  \frac{ 2(1+\eta\delta_{4k}^2)}{1-\delta^2_{4k }}  }+  \frac{\eta}{ 1-\delta_{4k}}\\
 & =  \sqrt{  \frac{ 2+ (\sqrt{5}+1) \delta_{4k}^2}{1-\delta^2_{4k }}  }  +  \frac{\sqrt{5}+1}{2(1-\delta_{4k})} .
    \end{align*}
We now prove that if $ \delta_{4k} $ satisfies (\ref{DDDLLL}), then the constant $ \rho <1. $
In fact, to ensure  $ \rho <1, $ it is sufficient to ensure that
$$  \delta_{4k}  \sqrt{  \frac{ 2(1+ \eta \delta_{4k}^2)} {1-\delta^2_{4k }}  }  <1, $$
 which, by squaring both sides,  can be written as
 $$ 2\eta \delta_{4k}^4 + 3 \delta_{4k}^2-1 <0.  $$
Thus
$$    \delta_{4k} ^2 < \frac{\sqrt{9+4(\sqrt{5}+1)}-3}{2(\sqrt{5}+1)}= \frac{2}{\sqrt{9+4(\sqrt{5}+1)}+3 }, $$
and hence
$$ \delta_{4k} <   \sqrt{\frac{2}{\sqrt{13+4\sqrt{5}}+3}}
 \approx 0.5102.  $$
The proof is complete. \hfill $\Box$

\vskip 0.1in

\begin{Rem}   \label{Rem-CoSaMP}
The  existing bound  $ \delta_{4k} <   0.4782 $ for CoSaMP  was shown by Foucart and Rauhut (see, e.g.,  Theorem 6.18 in  \cite{FR13}). Our result improves their result to $ \delta_{4k} <   0.5102. $  In terms of geometric rate $ 0.5$, Foucart and Rauhut's result  is equivalent to  $ \delta_{4k} <   0.299, $ and their bound was slightly improved to $ \delta_{4k} <   0.301   $ by Shen and Li \cite{SL18}.   Let us find out our RIP-based bound if the geometric  rate $\rho \leq 0.5  $ is required. To see this, let
 \begin{equation}  \label{0505}
  \rho =   \delta_{4k} \sqrt{  \frac{ 2( 1+  \eta \delta_{4k}^2) }{1-\delta^2_{4k }}  }  \leq 0.5.
  \end{equation}
% the inequality above is equivalent to the following univariate polynomial inequality:
%\begin{align*} &   4\eta^2 \delta^6_{4k} - 4(2\eta-1)\delta^5_{4k} -2(1+4\eta) \delta^4_{4k}+ (8\eta-17) \delta^3_{4k} \\
% & ~~ + 3 \delta^2_{4k} +12 \delta_{4k} -4 <0,
% \end{align*}
% where $ \eta = (\sqrt{5}+1)/2.$
 It is not difficult to verify that (\ref{0505}) is satisfied if  $$ \delta_{4k} \leq \sqrt{ \frac{2}{\sqrt{81+16(\sqrt{5}+1)}+9 } } \approx  0.3122, $$
 It guarantees the following estimation:
\begin{equation} \label{GR0.5} \|x^p-x_S\|_2 \leq 0.5^p \| x^0- x_S\|_2+ \gamma \| A^T  \nu '\|_2
 \end{equation}  where $ \gamma $ is a certain univariate constant. Clearly,  our result $\delta_{4k} \leq 0.3122$ for CoSaMP in terms of geometric rate $0.5$ also  improves  the existing result  established by Shen and Li \cite{SL18}.
\end{Rem}

\vskip 0.1in

\section{Conclusions}
The RIP-based bounds that guarantee the success of signal recovery  via the IHT and CoSaMP algorithms have been improved in this paper. A common feature of the two algorithms is using hard thresholding operator to produce a sparse approximation of the unknown signal. The property of  hard thresholding operator given in Lemma \ref{HHB} provides a useful basis to  the improvement of  the current performance theory for these algorithms.
The new   RIP-based bound $\delta_{2k} < (\sqrt{5}-1)/2 $  shown for IHT in this paper remarkably improves the current best known bound for this algorithm. While the improvement of the performance result for CoSaMP is much more challenging, the existing RIP-based bound for this algorithm was improved to $ \delta_{4k} < 0.5102 $  in this paper.    However, the question concerning the optimal (or the tightest) RIP-based bounds for the guaranteed performance of these algorithms  remains open at the moment.

 \end{document}